\documentclass[12pt]{article}
\input{epsf}

\def\lrpartial{\stackrel{\leftrightarrow}{\partial}}
\def\lrdiff{\stackrel{\leftrightarrow}{d}}

\begin{document}

\begin{titlepage}
\begin{flushright}
NSF-ITP-02-170\\
ITEP-TH-17/02\\
\end{flushright}

\begin{center}
{\Large $ $ \\ $ $ \\
Special contact Wilson loops}\\
\bigskip\bigskip
{\large Andrei Mikhailov\footnote{e-mail: andrei@kitp.ucsb.edu}}
\\
\bigskip\bigskip
{\it Kavli Institute for Theoretical Physics, University of California\\
Santa Barbara, CA 93106, USA\\
\bigskip
and\\
\bigskip
Institute for Theoretical and 
Experimental Physics, \\
117259, Bol. Cheremushkinskaya, 25, 
Moscow, Russia}\\

\vskip 1cm
\end{center}

\begin{abstract}
Wilson loops in ${\cal N}=4$ supersymmetric Yang-Mills theory
correspond at strong coupling to extremal surfaces
in $AdS_5$. We study a class of extremal surfaces
known as special Legendrian submanifolds. 
 The "hemisphere" corresponding to the
circular Wilson loop is an example of a special Legendrian
submanifold, and we give another example. 
We formulate the necessary conditions
for the contour on the boundary of $AdS_5$ to be the boundary of
the special Legendrian submanifold and conjecture that these
conditions are in fact sufficient. We call the solutions of these
conditions "special contact Wilson loops". The first order equations
for the special Legendrian submanifold impose a constraint on 
the functional derivatives of the Wilson loop at the special contact 
contour which should be satisfied in the Yang-Mills theory at 
strong coupling.
\end{abstract}
\end{titlepage}

\section{Introduction.}
Our understanding of the correspondence between gauge fields and strings
improved recently due to the development of the idea of 
the AdS/CFT correspondence \cite{Maldacena,GKP,Witten}. 
The  most remarkable achievement was the
comparison of the superstring and field theory computations of the
quantities which are not protected by the symmetries. 
An important example is the study of the Wilson loop 
functional \cite{ReyYee,MaldacenaWilson,ReyTheisenYee,BCFM,DGO}.
For the loop of the circular shape it was computed to all 
orders in the perturbation theory in the large $N$ Yang-Mills 
\cite{ESZ,DG}.
The result analytically continued to the strong coupling limit
was found to be in agreement with the string theory computation.
In another very recent development, the anomalous dimension
of the twist two operators in the ${\cal N}=4$ Yang-Mills
theory was computed from the properties of the extremal surfaces
\cite{Kruczenski,Makeenko}. Already these two examples show the 
importance of the study of the extremal surfaces in AdS space
for understanding the relation between gauge fields and strings.

In our paper we will study a class of the Wilson loops which corresponds
to a special class of the extremal surfaces in AdS --- special Legendrian
submanifolds. This is a large class of extremal surfaces which are easier
to study than the generic extemal surface. The reason for simplifications
is  that the special Legendrian submanifolds satisfy
the first order differential equations while the generic minimal
surfaces satisfy the second order differential equations.
We were not able to find explicitly the generic special Legendrian surface 
in AdS space and we think that it is actually not possible. 
But nontrivial explicit examples of the special Legendrian manifolds are known 
for $S^5$ \cite{CastroUrbano,Haskins,MarkGross,Joyce} and presumably can be 
constructed by similar
methods in $AdS_5$. In this paper we will consider only a simplest
nontrivial example and will mostly concentrate on general aspects
of the special Legendrian manifolds in $AdS_5$.

We will give the definition of the special Legendrian manifold
in Section 2. The surface ending on the circular contour on the boundary
which was found in \cite{BCFM,DGO} is a special case of a special Legendrian
manifold.  We will give another example in Section 2. In Section 3
we will find the necessary conditions for the contour in ${\bf R}\times S^3$
to be the boundary of the special Legendrian manifold in $AdS_5$ and find
all solutions to these conditions in terms of a real function of one real
variable.  We conjecture that at least for
the contours which are close to the circular contour these conditions
are in fact necessary and sufficient. We call such contours the
"special contact Wilson loops".
In Section 4 we will consider the infinitesimal deformations of the
special Legendrian manifold ending on the circular Wilson loop.
We will confirm by the explicit calculation that the infinitesimal
deformations  preserve 
the differential conditions on the special contact Wilson loop. 
In Section 5 we will show that the special contact Wilson loop
does not in general preserve any supersymmetry.
In Section 6 we will study the behavior of a special Legendrian manifold
near the boundary of the AdS space. In Section 7 we study 
the regularized area of the special Legendrian manifold.
We did not succeed in calculating the regularized area
for the general special contact boundary. This would presumably
require the knowledge of the actual special Legendrian surface.
But we do know something about the infinitesimal variation
of the regularized area under the variation of the contour (not
necessarily preserving the special contact condition). 
There is a special vector field $\lambda^{\mu}$ 
in ${\bf R}\times S^3$ which enters into 
the definition of the special Legendrian manifold. We show that
the variation of the regularized area functional under the 
infinitesimal deformation of the special contact contour is zero
provided that the normal vector describing the variation
is orthogonal to $\lambda^{\mu}$ pointwise on the contour. 
(The deformed contour does not have to be special contact.)
In other words ${\delta\over\delta C^{\mu}} W[C]\sim \lambda_{\mu}$ 
for
the special contact $C$. It would be interesting to see whether
this is true only at strong coupling.

An interesting feature of the special Legendrian surfaces is that the
worldsheet coordinates satisfy the first order differential equations.
This is usually associated with the supersymmetry. However the 
string wrapped on the special Legendrian manifold does not in fact
preserve any supersymmetry. A special case of the special contact
Wilson loop is the circular Wilson loop. On the field theory side
the circular Wilson loop classically preserves half of the 
supersymmetry. But quantum mechanically all the supersymmetry is
presumably broken 
because of the problems with the regularization \cite{BGK}.
 On the string
theory side the corresponding string worldsheet is not supersymmetric even
on the classical level. And for the special contact Wilson loops
which are not circular we show in Section 6 that 
the supersymmetry is completely broken already
in the field theory even classically.

\section{Special Legendrian manifolds.}
\subsection{Definitions.}
We consider $AdS_5$ embedded into ${\bf R}^{2+4}$ as the hyperboloid
\begin{equation}
y_{-1}^2+y_0^2=1+y_1^2+y_2^2+y_3^2+y_4^2
\end{equation}
The boundary of $AdS_5$ is the projectivization of the lightcone
$C$:
\begin{equation}
C:\;\;
y_{-1}^2+y_0^2=y_1^2+y_2^2+y_3^2+y_4^2
\end{equation}
The lightcone separates the future ${\bf R}^{2+4}_+$ from
the past ${\bf R}^{2+4}_-$.

We will introduce in ${\bf R}^{2+4}$ the complex coordinates
$z_0=y_{-1}+iy_0$, $z_1=y_1+iy_2$, $z_2=y_3+iy_4$.
The metric and the complex structure define the Kahler form
\begin{equation}
\omega=dz_0\wedge d\overline{z}_0-dz_1\wedge d\overline{z}_1
-dz_2\wedge d\overline{z}_2
\end{equation}
We will also need the holomorphic 3-form
\begin{equation}
\Omega=dz_0\wedge dz_1\wedge dz_2
\end{equation}
{\em Lagrangian submanifolds.}
The submanifold $L\subset {\bf C}^{1+2}$ is called Lagrangian
if $\omega|_{TL}=0$. One can see that for the Lagrangian manifold
\begin{equation}
\Omega|_{T_xL}=e^{i\phi(x)}\mbox{vol}_L
\end{equation}
{\em Special Lagrangian submanifolds.}
The Lagrangian manifold is called special Lagrangian 
\cite{HarveyLawson} if the phase $\phi(x)$
is constant (does not depend on $x$). In flat space 
it is enough to consider
the case when $\phi=0$ (because the cases with $\phi\neq 0$ 
are related to the cases with $\phi=0$ by a symmetry):
\begin{equation}
\mbox{Im}\; \Omega|_{TL}=0
\end{equation}
The special Lagrangian manifolds are extremal in the sense that 
when we deform them the variation of the volume is of the
second order in the deformation. We will review the
proof of this fact in Appendix.

The Euler vector field $E=x^{\mu}{\partial\over\partial x^{\mu}}$
is orthogonal to $AdS_5$. The complex structure in ${\bf R}^{2+4}$
defines the one-form $\lambda$ in the AdS space:
\begin{equation}
\lambda=\iota_{E}\omega
\end{equation}
The corresponding vector field $\lambda^{\mu}$ can be restricted to the
boundary; the restriction is also denoted $\lambda^{\mu}$.
The restriction of the Kahler form $\omega$ to the AdS space is
$\omega|_{AdS}={1\over 2}d\lambda$.
\vspace{10pt}

\noindent
{\em Contact submanifolds.} 
A submanifold $X$ of  $AdS_5$ is called contact if the restriction
of $\lambda$ on $X$ is zero. This is equivalent to the cone over $X$ being
Lagrangian. The maximal dimension of a contact submanifold  $X$ in $AdS_5$ 
is two.   
\vspace{10pt}

\noindent
{\em Special Legendrian submanifolds.} 
 A special Lagrangian manifold is called
a special Lagrangian cone if it is invariant under the rescalings
generated by $E$. The intersection of a special Lagrangian
cone with $AdS_5$ is called a special Legendrian submanifold
\cite{Haskins}. A special Legendrian submanifold is contact and the
restriction of $\epsilon^{\lambda\mu\nu}z_{\lambda} dz_{\mu}\wedge 
dz_{\nu}$ to its tangent space is real. 
It is an extremum of the area functional. 
\vspace{10pt}

\noindent
{\em The "hemisphere".}
An example of the special Lagrangian cone is the plane given by the
equations 
\begin{equation}\label{plane}
y_0=y_1=y_4=0
\end{equation}
The corresponding special Legendrian manifold is $AdS_2\subset AdS_5$.
Its boundary is the circular Wilson loop. Historically it was one of the first
Wilson loops computed in the strong coupling limit by the
AdS/CFT correspondence \cite{BCFM,DG}. The extremal surface
looks like a hemisphere in the Poincare coordinates. 

The infinitesimal deformations of the Lagrangian submanifold $\Sigma$ are in
one to one correspondence with the generating functions $H$; the corresponding
normal vector field $\xi_H$ is given by the equation
\begin{equation}
dH(\eta) = \omega(\xi_H,\eta)
\end{equation}
for any vector $\eta$ tangent to $\Sigma$. If we require $\Sigma$ to be a cone
$H$ has to satisfy $H(ty)=t^2H(y)$ --- a homogeneous function of the degree 2. 
If we require both $\Sigma$ and its deformation
 to be special Lagrangian then $H$ has to be harmonic:
\begin{equation}
{\partial^2 H\over\partial y^{\mu}\partial y_{\mu}}=0
\end{equation}
Therefore the deformations of the special Lagrangian cones are parametrized
by a single homogeneous harmonic function of the degree two.
In some sense there are as many special Lagrangian cones as there
are harmonic functions of the degree two on a three-dimensional space. 
Notice that
the deformations of the general three-dimensional extremal cone in ${\bf R}^{2+4}$
are parametrized roughly speaking by three homogeneous harmonic functions of
the degree two (the worldsheet coordinates).   
The general extremal cone is parametrized by three harmonic
functions while the special Lagrangian cone is parametrized by one
harmonic function.

\subsection{Special Legendrian manifolds are described by the first order
differential equations.}
The special Legendrian manifolds have a property which resembles the
principle of analytic continuation for the complex curves.
It turns out that the special Legendrian manifold is completely
determined by any one-dimensional contour belonging to it 
\cite{HarveyLawson}. 
Consider the contour $z_{\mu}(\sigma)$, $\sigma\in {\bf R}$ inside
the special Legendrian manifold $\Sigma$. At any given point 
on the contour $z_{\mu}(\sigma_0)$ the tangent space to $\Sigma$
is generated by two vectors one of which is 
$\partial_{\sigma}z_{\mu}(\sigma_0)$. The other one may be chosen to be
orthogonal to $\partial_{\sigma}z_{\mu}$. Let us call it $\xi$.
It turns out that $\xi$ is completely determined up to the multiplication
by a real number by the condition that $\Sigma$ is special Legendrian.
Indeed, $\xi$ should satisfy the equations:
\begin{equation}
\left\{
\begin{array}{l}
z^*_{\mu}\xi^{\mu}=0\\[5pt]
\partial_{\sigma}z^*_{\mu}\xi^{\mu}=0\\[5pt]
\epsilon_{\mu\nu\lambda}z^{\mu}\partial_{\sigma} z^{\nu}
\xi^{\lambda}\in {\bf R}
\end{array}
\right.
\end{equation}
The first of these equations says that $\xi^{\mu}$ belongs
to $AdS_5$ (real part) and to the kernel of $\lambda$ (imaginary
part). The second equation says that $\xi^{\mu}$ is orthogonal
to $\partial_{\sigma}z$ (real part) and 
$\omega(\partial_{\sigma}z,\xi)=0$
(imaginary part).
These are five real equations on six real components of $\xi^{\mu}$
therefore the direction of $\xi^{\mu}$ is determined:
\begin{equation}\label{DirectionXi}
\xi^{\mu}(\sigma)\sim\epsilon^{\mu\nu\lambda}\overline{z}_{\nu}
\lrpartial\overline{z}_{\lambda}
\end{equation}
We can deform the contour 
$z^{\mu}(\sigma)\to z^{\mu}(\sigma)+\epsilon \xi^{\mu}(\sigma)$.
The deformed contour still belongs to $\Sigma$. Therefore
we get a family of contours sweeping $\Sigma$. This family
can be parametrized by a real parameter $\tau$:
\begin{equation}\label{FirstOrder}
\begin{array}{l}
\partial_{\tau}z_0=\overline{z}_1\lrpartial_{\sigma} \overline{z}_2 \\[5pt]
\partial_{\tau}z_1=-\overline{z}_2\lrpartial_{\sigma}\overline{z}_0 \\[5pt]
\partial_{\tau}z_2=\overline{z}_1\lrpartial_{\sigma}\overline{z}_0
\end{array}
\end{equation}
This is a system of the first order equations on the worldsheet coordinates.

By this construction any contact one-dimensional contour in $AdS_5$ will give
a special Legendrian manifold. Contact one-dimensional contours in $AdS_5$
depend on three real function of a real variable. Two one-dimensional
contours give the same special Legendrian manifold if they are
related by the deformation (\ref{DirectionXi}). Therefore 
a special Legendrian submanifold in $AdS_5$ is parametrized by two real
functions of a real variable. This probably suggests that a special
Legendrian manifold will generally have at least two boundaries. 
Indeed, as we have seen these manifolds are parametrized roughly
speaking by a harmonic function; but a harmonic function is defined
by its boundary values which gives one real function of a real variable per
boundary.

\subsection{Example.}
In this section we will repeat in $AdS_5$ the construction
of the special Legendrian manifold in $S^5$ suggested in 
\cite{Joyce}.
We consider the following surface parametrized
by the two real parameters $\sigma$ and $\tau$:
\begin{equation}
z_{\mu}(\tau,\sigma)=ig_{\mu}(\tau) e^{i\alpha_{\mu} \sigma}
\end{equation}
where $g_{\mu}(\tau)$ is real. 
This is a special Legendrian manifold if
\begin{equation}\label{SumAlpha}
\alpha_0+\alpha_1+\alpha_2=0
\end{equation}
and  $g_0(\tau)$, $g_(\tau)$, $g_2(\tau)$ satisfy the algebraic equations:
\begin{equation}\label{AlgebraicForG}
\begin{array}{l}
g_0(\tau)^2-g_1(\tau)^2-g_2(\tau)^2=1\\[5pt]
\alpha_0 g_0(\tau)^2-\alpha_1 g_1(\tau)^2- \alpha_2 g_2(\tau)^2=0\\[5pt]
\end{array}
\end{equation}
Indeed $z_{\mu}(\tau,\sigma)$ satisfies (\ref{FirstOrder}) if
we choose $\tau$ so that $g_{\mu}(\tau)$ satisfy the system of differential
equations:
\begin{equation}
\begin{array}{l}
\dot{g}_0=-(\alpha_1-\alpha_2)g_1g_2\\[5pt]
\dot{g}_1=(\alpha_2-\alpha_0)g_2g_0\\[5pt]
\dot{g}_2=-(\alpha_1-\alpha_0)g_1g_0
\end{array}
\end{equation}
for which (\ref{AlgebraicForG}) are integrals of motion.
Equations (\ref{AlgebraicForG}) have real solutions only if
$\alpha_1$ and $\alpha_2$ have different sign.
Also notice that $\alpha_{\mu}\mapsto -\alpha_{\mu}$ is a symmetry.
This means that without any loss of generality we may assume
\begin{equation}
-\alpha_2>\alpha_1>0
\end{equation}
With this choice of $\alpha_{\mu}$ the solution is:
\begin{equation}\label{Solution}
\begin{array}{l}
z_0= i\left((\alpha_1-\alpha_2)T+{1\over 3}\right)^{1/2}
e^{-i(\alpha_1+\alpha_2)\sigma}
\\[5pt]
z_1= i\left((-\alpha_1-2\alpha_2)T-{1\over 3}\right)^{1/2}
e^{i\alpha_1\sigma}\\[5pt]
z_2= \pm i\left((2\alpha_1+\alpha_2)T-{1\over 3}\right)^{1/2}
e^{i\alpha_2\sigma}
\end{array}
\end{equation}
where $T\in\left[{1\over 3(2\alpha_1+\alpha_2)}, +\infty\right]$.
The induced metric on the worldsheet is:
\begin{equation}
\begin{array}{l}
ds^2=-\left[T(\alpha_1-\alpha_2)(-\alpha_1-2\alpha_2)(2\alpha_1+\alpha_2)-
{2\over 3}(\alpha_1^2+\alpha_1\alpha_2+\alpha_2^2)\right]\times
\\[5pt]
\times\left\{ d\sigma^2
+ \left(\left[(\alpha_1-\alpha_2)T+{1\over 3}\right]
\left[(-\alpha_1-2\alpha_2)T-{1\over 3}\right]
\left[(2\alpha_1+\alpha_2)T-{1\over 3}\right]\right)^{-1}{dT^2\over 4}
\right\}
\end{array}
\end{equation}
This metric is negative definite and becomes asymptotically the metric of 
the $AdS_2$ when $T\to +\infty$.
The Laplacian of the worldsheet coordinates in the induced metric is:
\begin{equation}
\begin{array}{l}
{1\over 4}
\left[(\alpha_1-\alpha_2)T+{1\over 3}\right]^{-{1\over 2}}
\left[(-\alpha_1-2\alpha_2)T-{1\over 3}\right]^{-{1\over 2}}
\left[(2\alpha_1+\alpha_2)T-{1\over 3}\right]^{-{1\over 2}}
{\partial^2 z_{\mu}\over\partial\sigma^2}+\\[5pt]
+{\partial\over\partial T}
\left[(\alpha_1-\alpha_2)T+{1\over 3}\right]^{1\over 2}
\left[(-\alpha_1-2\alpha_2)T-{1\over 3}\right]^{1\over 2}
\left[(2\alpha_1+\alpha_2)T-{1\over 3}\right]^{1\over 2}
{\partial z_{\mu}\over\partial T}=\\[10pt]=
{1\over 2}
{T(\alpha_1-\alpha_2)(-\alpha_1-2\alpha_2)(2\alpha_1+\alpha_2)-
{2\over 3}(\alpha_1^2+\alpha_1\alpha_2+\alpha_2^2)\over
\sqrt{\left[(\alpha_1-\alpha_2)T+{1\over 3}\right]
\left[(-\alpha_1-2\alpha_2)T-{1\over 3}\right]
\left[(2\alpha_1+\alpha_2)T-{1\over 3}\right]}} z_{\mu}
\end{array}
\end{equation}
which explicitly shows that the surface is extremal.
The boundary is at $T=+\infty$. It can be parametrized by $\sigma$:
\begin{equation}
\begin{array}{l}
z_0=i\sqrt{\alpha_1-\alpha_2}\; e^{-i(\alpha_1+\alpha_2)\sigma}\\[5pt]
z_1=i\sqrt{-\alpha_1-2\alpha_2}\; e^{i\alpha_1 \sigma}\\[5pt]
z_2=\pm i\sqrt{\alpha_2+2\alpha_1}\; e^{i\alpha_2\sigma}
\end{array}
\end{equation} 
It is spacelike and 
consists of two components corresponding to the choice of the sign
in the formula for $z_2$. It is interesting to consider the limit
$\alpha_2=-2\alpha_1+a$ where $a$ is positive and small. The boundary
consists of two spirals with the common central line. It is spacelike
becoming lightlike when $a=0$. The distance between the spirals is
comparable to the length of the period of each spiral. 

An interesting property of this example is that the double spiral
extremizes the Wilson loop functional. We will prove it
at the end of Section 7.

\section{Special contact Wilson loops.}
We want to characterize the one-dimensional contours in the boundary of 
$AdS_5$ which are the boundaries of the special Legendrian manifolds
in $AdS_5$. We will first describe some necessary conditions for such a 
contour. The first condition is that the contour
should be contact. This means that the one form on the boundary
of the AdS space which defines the contact structure should be
zero on a tangent vector to the contour. This condition by itself
is not enough. To formulate the second condition we define another
one-form on the boundary. More precisely, this one-form is defined
only on those vectors which are tangent to the contact structure.
We prove that if the contour is the boundary of the special 
Lagrangian cone then this second one-form should be also zero 
on the contour.

We then conjecture that these two necessary conditions are sufficient
at least for those contours which are small deformations of a 
circular contour. We explicitly describe the solutions to these
conditions. 

\subsection{Two necessary conditions.}
The boundary of $AdS_5$ is conformally $S^1\times S^3$.
It is the projectivization of the
lightcone $C\subset {\bf R}^{2+4}$.
Instead of considering the one-dimensional curves in $S^1\times S^3$
we will consider the two-dimensional subcones of $C$.
Suppose that $X\subset C$ is a two-dimensional subcone.
When $X$ is the boundary of the special Lagrangian cone
$L\subset {\bf R}^{2+4}_+$?
The obvious necessary condition is that the restriction of $\omega$
on $X$ is zero:
\begin{equation}\label{FirstCondition}
\iota_E\omega=
z^*_{\mu}\stackrel{\leftrightarrow}{d}z^{\mu}=0
\end{equation}
To formulate the second necessary condition we will need to define
a complex one-form $\Lambda$.
Let $TX$ be the tangent bundle to $X$. Let
$S\subset TX$ be the subspace tangent to the contact structure:
\begin{equation}
S=\{\xi\in TX | \omega(E,\xi)=0\}
\end{equation}
Consider the following one-form on $S$:
\begin{equation}
\Lambda=\left.{z_1 dz_2- z_2 dz_1\over z_0^*}\right|_{S}
\end{equation}
Notice that $\Lambda$ is $SU(1,2)$ invariant. Indeed one can see
that
\begin{equation}\label{ThreeFormsEqual}
 \left.{z_1 dz_2- z_2 dz_1\over z_0^*}\right|_{S}
=\left.{z_0 dz_2- z_2 dz_0\over z_1^*}\right|_{S}
=\left.{z_1 dz_0- z_0 dz_1\over z_2^*}\right|_{S}
\end{equation}
These three forms of rewriting $\Lambda$ prove the invariance
of $\Lambda$ under $SU(0,2)$ which stabilizes $z_0$ and
two $SU(1,1)$ which stabilize $z_1$ and $z_2$. These three groups
generate $SU(1,2)$.

For $X$ to be the boundary of the special Lagrangian $L$ it is
necessary that
\begin{equation}\label{SecondCondition}
\mbox{Im}\;\Lambda|_{TX}=0
\end{equation}
Let us prove that this is a necessary condition.

Let us fix a point $l\in X$. The tangent space $T_lX$ is
generated by two linearly independent vectors $l$ and $\eta$,
where $\eta$ should be orthogonal to $l$.
To generate the tangent space to $L$ we need to add a
third vector $\nu$ which is orthogonal to $\eta$ and
leads out of the light cone.
By an $SU(1,2)$ transformation we can bring $\eta$ to the form
\begin{equation}
\eta=\left[\begin{array}{c} 0\\ 0\\ \eta_2 \end{array}\right],\;\;
\eta_2\in {\bf C}
\end{equation}
We know that $l$ is orthogonal to $\eta$ and $I.\eta$. Therefore
$l$ should be of the form:
\begin{equation}
l=\left[\begin{array}{c}q_0\\ q_1\\ 0\end{array}\right]
\end{equation}
Consider the subgroup $U(1,1)\subset SU(1,2)$ which rotates 
$\eta$ by a phase: 
$U(1,1)=\{g\in SU(1,2)\;|\; g.\eta= e^{i\phi}\eta\}$.
We can use this subgroup to make $q_0$ and $q_1$ real. Notice 
that $q_0^2- q_1^2=0$. What can we say about $\nu$?
Since $L$ is special Lagrangian $\nu$ should be orthogonal to
both $I.\eta$ and $I.l$. Also remember that we have chosen
$\nu$ to be orthogonal to $\eta$. This means that $\nu$ is
of the form:
\begin{equation}
\nu = \left[\begin{array}{c}
r_0\\
r_1\\
0
\end{array}\right]\;+\; a\; I.l
\end{equation}
where $r_0,r_1$ and $a$ are real numbers.
Now we have
$\Omega(\nu,\eta,l)=(r_0 q_1- r_1 q_0)\eta_2$ and
$\Lambda(\eta)={q_1\eta_2\over q_0}$. Therefore
$\Omega(\nu,\eta,l)\in {\bf R}$ implies $\Lambda(\eta)\in{\bf R}$.
This proves that (\ref{SecondCondition}) is a necessary condition.

We conjecture  that at least for the contours sufficiently
close to the circle (\ref{FirstCondition}) and (\ref{SecondCondition})
are actually  sufficient.  Our motivation for
this conjecture is the counting of the parameters. 
Consider the "hemisphere" (\ref{plane}). The small deformations of this 
hemisphere correspond to the degree two harmonic functions on 
the corresponding cone. These harmonic functions should be determined
by their boundary values (we will explain the details in Section 4.)
The boundary value of the function is a real function on a circle.
We conjecture that these deformations are unobstructed.  This was
proven in \cite{McLean} for compact special Lagrangian manifolds,
but we are dealing with non-compact cases. If it is true that the
deformations corresponding to harmonic functions are unobstructed
then  the special Legendrian manifolds close to the
hemisphere should be parametrized by a real function on a circle.
In the next subsection we will see that the contours satisfying
(\ref{FirstCondition}) and (\ref{SecondCondition}) are also 
parametrized by a real function on a circle. This suggests that
there is a one to one correspondence between the contours
satisfying (\ref{FirstCondition}) and (\ref{SecondCondition}) and
the special Legendrian manifolds, at least in the vicinity 
of the circular contour.

\subsection{Special contact loops.}
We will call special contact loops the solutions to the necessary
conditions (\ref{FirstCondition}) and (\ref{SecondCondition}):
\begin{equation}\label{Conditions}
\begin{array}{l}
\iota_E\omega (\eta)=0\\[5pt]
\mbox{Im}\;\Lambda (\eta)=0
\end{array}
\end{equation}
These special contact loops can be described very explicitly.
Consider a closed path
 $u(\sigma)$ in $S^2=CP^1$ which restricts the domain
of zero area (area is counted with the orientation; 
$\infty$ is an example of a path which restricts
zero area.)
 This path can be lifted to the horizontal curve in
$S^3\stackrel{S^1}{\rightarrow}CP^1$ which satisfies
\begin{equation}
y_1^*\stackrel{\leftrightarrow}{\partial_{\sigma}} y_1+
y_2^*\stackrel{\leftrightarrow}{\partial_{\sigma}} y_2=0,\;\;\;
y_2(\sigma)/y_1(\sigma)=u(\sigma)
\end{equation}
This gives a solution to (\ref{Conditions}):
\begin{eqnarray}
&&z_1(\sigma)=
e^{i\psi(\sigma)}y_1(\sigma)\\[5pt]&&
z_2(\sigma)=e^{i\psi(\sigma)}
y_2(\sigma)\\[5pt]&&
z_0(\sigma)=e^{i\psi(\sigma)}
\sqrt{|y_1(\sigma)|^2+|y_2(\sigma)|^2}
\end{eqnarray}
where
\begin{equation}
e^{i\psi(\sigma)}=
\left({y_1^*(\sigma)\lrpartial_{\sigma} y_2^*(\sigma)\over
y_1(\sigma)\lrpartial_{\sigma} y_2(\sigma)}\right)^{1\over 6}
\end{equation}
Therefore the special contact Wilson loops correspond to the
closed contours in $CP^1$ restricting a domain of zero area.

\subsection{In Poincare coordinates.}
Let us introduce the Poincare coordinates:
\begin{equation}
(z_0;z_1,z_2)=\left({1-x_{\mu}^2+h^2\over 2h}+i{x_0\over h};
{x_1+ix_2\over h}, {x_3\over h}+i{1+x_{\mu}^2-h^2\over 2h}\right)
\end{equation}
Let us write the conditions for the contour 
in these coordinates. It is convenient to
introduce: 
\begin{equation}
x_{\pm}=x_0\pm x_3
\end{equation}
 The contact condition $\lambda=0$ reads:
\begin{equation}\label{LZeroPoincare}
d\;x_+ + x_- \stackrel{\leftrightarrow}{d} x_{\mu}^2
-2 x_1\stackrel{\leftrightarrow}{d} x_2 =0
\end{equation}
We  find it   more convenient to consider  special
Lagrangian manifolds with $i\Omega\in {\bf R}$
(rather than  $\Omega\in{\bf R}$)
 when working in Poincare coordinates. 
The special condition 
$\mbox{Re}[z_0(z_1\stackrel{\leftrightarrow}{d}z_2)]=0$ becomes:
\begin{equation}\label{SpecialPoincare}
\mbox{Re}\; \left[
(x_1+ix_2)\left( {1\over 2}dx_+-
{1\over 2}x_-\lrdiff x_{\mu}^2+ ix_0\lrdiff x_3 +
{i\over 2} d\;x_{\mu}^2\right)\right]=0
\end{equation}
The solutions to these conditions are parametrized by a
a complex valued function $y(\sigma)$:
\begin{equation}\label{SCPoincare}
\begin{array}{rcl}
x_1+ix_2&=&y(\partial_{\sigma}\bar{y})^{1/3}/
\mbox{Re}\;(\partial_{\sigma}y)^{1/3}
\\[5pt]
x_-&=&-\mbox{Im}\;(\partial_{\sigma}y)^{1/3}/
     \mbox{Re}\;(\partial_{\sigma}y)^{1/3}
\\[5pt]
x_+&=&x_-|y|^2+i\int d\sigma\;(y\partial_{\sigma}\bar{y}-
\bar{y}\partial_{\sigma}y)+C
\end{array}
\end{equation}
where $C$ is a constant. Adding constant to $x_+$ corresponds
to the $su(1,2)$ transformation $\delta(z_0;z_1,z_2)=
(iz_0+z_2;0,z_0-iz_2)$.

\section{Infinitesimal 
deformations of the special Lagrangian plane.}
As we have explained in Section 2 the deformations of the
special Lagrangian cone correspond to the harmonic functions,
homogeneous of the degree two. If the cone is a plane we
can describe such functions and the corresponding deformations rather
explicitly. 

An infinitesimal
deformation of the plane ${\bf R}^{1+2}\subset {\bf R}^{2+4}$ 
is described by a vector field $\xi(v)$, $v\in {\bf R}^{1+2}$ 
which is orthogonal to ${\bf R}^{1+2}$. 
The Lagrangian deformations correspond to functions $H$ on
${\bf R}^{1+2}$ in the following way:
$$\xi_H(v)=I.\nabla H(v)$$
where $I$ is the complex structure in ${\bf R}^{2+4}$ (multiplication
by $i$ in ${\bf C}^{1+2}$.) 
We want the deformed submanifold to be a special Lagrangian
cone. This leads us to considering the harmonic functions $H$
which are homogeneous, $H(tv)=t^2H(v)$.
Such a function $H(v)$ can be reconstructed 
from its values on the lightcone. 
Let us choose a vector 
$v_0=[1,0,0]$ in ${\bf R}^{1+2}$ and consider the circle
$S(v_0)$ --- the set of points $l$ on the lightcone  $(l,l)=0$
satisfying $(l,v)=1$.
According to the Asgeirsson theorem about the mean value of the harmonic
function\footnote{For the explanation of the Asgeirsson theorem
see for example \cite{CH}.},
\begin{equation}
H(v)={1\over 3\pi}\int_{S(v_0)} {dl H(l) v^5\over (v\cdot l)^3}
\end{equation}
In fact we could have replaced $S(v_0)$ by any closed path on the lightcone;
this integral with the naturally defined measure $dl$ does not depend
on the choice of the path. 
The limiting value near the
intersection with the light cone $\xi(v)|_{v\to l}$ depends
on $H(l)$, ${d\over d\sigma}H(l)$ and
${d^2\over d\sigma^2}H(l)$ where $\sigma$ is the angular coordinate
on $S(v_0)$.  The direct computation expressing
$\xi(v)|_{v\to l}$ through $H(l)$ and its first and second derivative 
should be rather cumbersome. We will use
a trick. First let us evaluate a particular integral:
\begin{equation}
I[Q](v)={1\over 3\pi}\int {v^5\over (v\cdot l)^3} (l,Q.l)[dl]
\end{equation}
where $Q$ is a constant $6\times 6$ matrix. Because of the
$SO(2,4)$ invariance
\begin{equation}
I[Q](v)=Av^2\mbox{tr}\;Q +B(v, Q.v)\\[5pt]
\end{equation}
From $I[Q={\bf 1}](v)=0$ we get
\begin{equation}
A=-{1\over 3}B
\end{equation}
From the theorem about the mean value, $B=1$. Therefore
\begin{equation}
I[Q](v)=-{1\over 3}v^2\mbox{tr}\;Q +(v,Q.v)
\end{equation}
Let us take
\begin{equation}
H_Q(l)=(l,Q.l)\;\;\mbox{with}\;\;
Q=\left[\begin{array}{ccc} Q_{00} &  Q_{01} &  Q_{02} \\[5pt]
                          -Q_{01} &  Q_{11} &  Q_{12} \\[5pt]
                          -Q_{02} &  Q_{12} &  Q_{22}
        \end{array}\right]
\end{equation}
Take $l(\sigma)=(1,\cos\sigma, \sin\sigma)$. 
We will write $H(\sigma)$ instead of $H(l(\sigma))$.
At $\sigma=0$ we have
$H_Q(0)=Q_{00}+2Q_{01}-Q_{11}$, $H_Q'(0)=2(Q_{02}-Q_{12})$,
$H_Q''(0)=2(Q_{11}-Q_{01}-Q_{22})$.  At the same time
\begin{equation}
\begin{array}{l}
\nabla I[Q]=-{2\over 3}v\;\mbox{tr}\; Q + 2 Q.v =\\[5pt]=
-{2\over 3}\times\left[\begin{array}{c}
Q_{00}+Q_{11}+Q_{22}-3(Q_{00}+Q_{01})\\[5pt]
Q_{00}+Q_{11}+Q_{22}-3(-Q_{01}+Q_{11})\\[5pt]
3(Q_{02}-Q_{12})
\end{array}\right]=
{1\over 3}\times
\left[\begin{array}{c} 
H_Q''(0)+4H_Q(0)\\
-2H_Q(0)+H_Q''(0)\\
-3H_Q'(0)
\end{array}\right]
\end{array}
\end{equation}
If the boundary data has $H(0)=H_Q(0)$, $H'(0)=H_Q'(0)$
and $H''(0)=H_Q''(0)$ then $\nabla H(0)=\nabla H_Q(0)$. 
Therefore:
\begin{equation}
\nabla H(\tau)={1\over 3}\times \left[\begin{array}{c}
H''(\tau)+4H(\tau)\\[5pt]
(H''(\tau)-2H(\tau))\cos\tau +3H'(\tau)\sin\tau\\[5pt]
-3H'(\tau)\cos\tau +(H''(\tau)-2H(\tau))\sin\tau
\end{array}\right]
\end{equation}
One can verify that this deformation preserves both 
$\bar{z}_{\mu}\lrdiff z^{\mu}=0$ and 
Im$[z_0(z_1\lrdiff z_2)]=0$. Indeed the unperturbed
contour is $(z_0;z_1,z_2)=(1;\cos\tau,\sin\tau)$. 
The contact form:
\begin{equation}
(I.(l+I.\nabla H),\partial_{\tau}l+I\partial_{\tau}\nabla H)=
-(\nabla H,\partial_{\tau}z)+(z,\partial_{\tau}\nabla H)=0
\end{equation}
The special condition:
\begin{equation}
\xi_0(\cos\tau\lrpartial_{\tau}\sin\tau)+
(\xi_1\lrpartial_{\tau}\sin\tau)+
(\cos\tau\lrpartial_{\tau}\xi_2)=0
\end{equation}
In this example we see that the deformation of the special
contact Wilson loop is described in terms of a single
function $H(\sigma)$. The formula for the deformation
is rather complicated involving up to two derivatives of $H$.
The deformation preserves the special contact conditions
(\ref{FirstCondition}) and (\ref{SecondCondition}).

\section{No supersymmetry.}
The extremal surface in $AdS_5$ with the boundary on the
contour $C$ corresponds on the field theory side to the
insertion of the Wilson loop functional:
\begin{equation}
W[C]={1\over N}\mbox{tr}\;P\exp\int d\sigma
(iA_{\mu}\partial_{\sigma}x^{\mu}+\Phi_1|\partial_{\sigma}x|)
\end{equation}
This functional is invariant under the superconformal
transformations which are generated by the conformal Killing
spinor $\psi(x)$ satisfying the constraint:
\begin{equation}\label{SUSY}
\gamma_{\mu}\partial_{\sigma}x^{\mu}\psi(x)=
i|\partial_{\sigma} x|\Gamma_1\psi(x)
\end{equation}
Here $\gamma_{\mu}$ are the space-time gamma matrices and
$\Gamma_1$ is first of the six gamma-matrices generating
the Clifford algebra of ${\bf R}^6$. We should have 
$\{\gamma_{\mu},\gamma_{\nu}\}=2g_{\mu\nu}$ and 
$\{\Gamma_i,\Gamma_j\}=2\delta_{ij}$ and 
$[\gamma_{\mu},\Gamma_i]=0$.
The conformal Killing spinors are of the form
\begin{equation}
\psi(x)=\psi_0+\gamma_{\mu}x_{\mu}\psi_1
\end{equation}
where both $\psi_0$ and $\psi_1$ are constant 
 spinors\footnote{The conformal Killing spinors on flat
${\bf R}^4$ should satisfy
$\partial_{\mu}\psi=\gamma_{\mu}\psi_1$. It follows that
$\psi_1$ should be constant. Indeed, $\psi_1$ would satisfy
$\gamma_{\mu}\partial_{\nu}\psi_1=\gamma_{\nu}\partial_{\mu}
\psi_1$ which implies that $\psi_1$ is a constant.}.
The condition (\ref{SUSY}) is satisfied for the circular Wilson loop. 
Indeed, consider the circular Wilson loop in the plane $(x_1,x_2)$ 
given by the equation:
\begin{equation}
x_1^2+x_2^2=1
\end{equation}
The condition (\ref{SUSY}) is satisfied for the following 
conformal Killing spinor:
\begin{equation}
\psi(x)=\chi_0-i(x_1\gamma_1+x_2\gamma_2)\gamma_1\gamma_2 \Gamma_1\chi_0
\end{equation}
where $\chi_0$ is an arbitrary constant spinor. 
But the generic special contact Wilson loop does
not preserve any supersymmetry. For example let us consider
the Wilson loop corresponding to the contour $y(\sigma)$ 
shown on the picture:
\begin{center}
\leavevmode
\epsffile{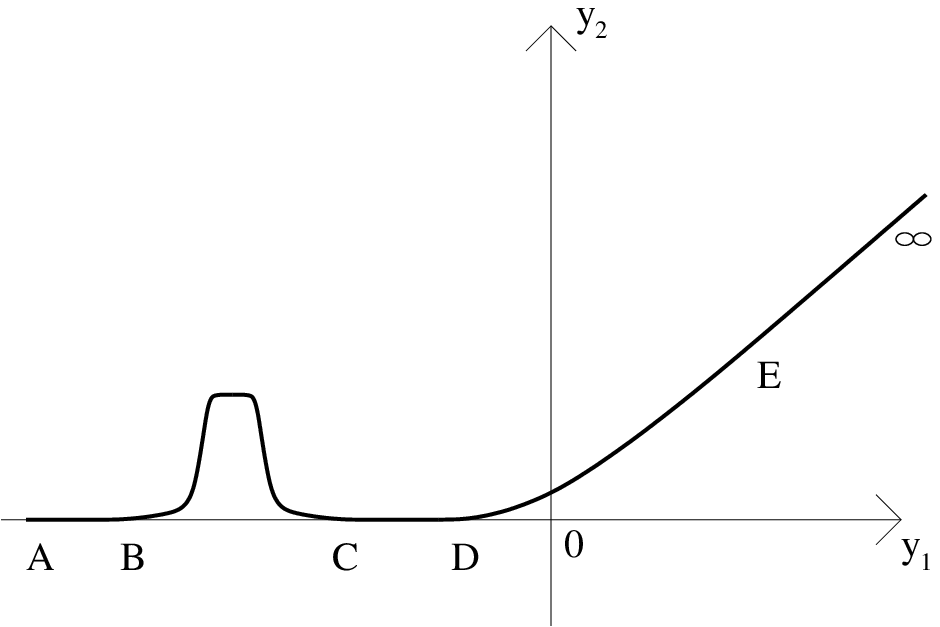}
\end{center}
Suppose that this contour preserves the conformal Killing
spinor $\psi_0+\gamma_{\mu}x_{\mu}\psi_1$. 
Let us consider three intervals $(A,B)$ $(C,D)$ and 
$(E,\infty)$. For the interval $(A,B)$:
\begin{equation}
\gamma_1(\psi_0+x_1\gamma_1\psi_1)=
i\Gamma_1(\psi_0+x_1\gamma_1\psi_1)
\end{equation}
which implies that $\gamma_1\psi_0=i\Gamma_1\psi_0$
and $\gamma_1\psi_1=i\Gamma_1\psi_1$. For the interval 
$(C,D)$:
\begin{equation}
\gamma_1(\psi_0+(x_1\gamma_1+x_+\gamma_-)\psi_1)=
i\Gamma_1(\psi_0+(x_1\gamma_1+x_+\gamma_-)\psi_1)
\end{equation}
therefore $\gamma_-\psi_1=0$. 
Finally, for the interval $(E,\infty)$
\begin{equation}
\begin{array}{l}
{1\over\sqrt{\dot{x}_1^2+\dot{x}_2^2}}
(\dot{x}_1\gamma_1+\dot{x}_2\gamma_2+
\dot{x}_+\gamma_-)
(\psi_0+(x_1\gamma_1+x_2\gamma_2+x_-\gamma_+)\psi_1)
=\\[5pt]=
i\Gamma_1(\psi_0+(x_1\gamma_1+x_2\gamma_2+x_-\gamma_+)
\psi_1)
\end{array}
\end{equation}
In the limit $|x|\to\infty$:
\begin{equation}
{1\over\sqrt{\dot{x}_1^2+\dot{x}_2^2}}\dot{x}_+x_-\psi_1=
i\Gamma_1(x_1\gamma_1+x_2\gamma_2)\psi_1
\end{equation}
This would imply that $\gamma_2\psi_1=\pm i \Gamma_1\psi_1$
which contradicts $\gamma_1\psi_1=i\Gamma_1\psi_1$.
Therefore the special contact loops are generally not
supersymmetric.

\section{Behavior near the boundary.}
In this section we will study the special Legendrian submanifold $X$
in the vicinity of a point on its boundary. The main result is
the following. Consider a curve inside $X$
which originates from the point $l_0$ on the boundary of $X$ and is orthogonal
to the boundary at this point. Consider the acceleration of this curve
and take  the component of the acceleration normal to $X$. It turns out
that the normal component of the acceleration is directed along 
$I.l_0$. We will first prove it in the simpler case when the contour
is locally exactly a straight line, and then give a general proof.
\vspace{10pt}

\noindent
{\em Special choice of the Poincare coordinates.}
Fix a point $l_0$ on the boundary of our special Legendrian submanifold.
We will use the Poincare coordinates:
\begin{equation}
(z_0;z_1,z_2)=\left({1-x_{\mu}^2+h^2\over 2h}+i{x_0\over h};
{x_1+ix_2\over h}, x_3+i{1+x_{\mu}^2-h^2\over 2h}\right)
\end{equation}
The point $l_0$  corresponds to $(1;0,i)$. 
We will choose the Poincare coordinates in such a way that the Wilson 
loop near the point $l_0$ is nearly a straight line:
\begin{equation}\label{CubicShape}
x_0=\beta_0 x_1^3+\ldots, \;\;\;
x_2=\beta_1 x_1^3+\ldots, \;\;\;
x_3=\beta_2 x_1^3+\ldots
\end{equation}
In other words the curvature of the Wilson loop at the point $l_0$ is zero.
Notice that we can always choose such coordinates. Indeed the Wilson
loop in the vicinity of $l_0$ is nearly a circle. This circle is
the boundary  of the intersection of the AdS space with some plane.
Because the Wilson loop is special contact the plane is special Lagrangian.
Let us choose another lightlike vector $\tilde{l}_0$ on the plane and
a space like vector $e$ 
on the plane orthogonal to both $l_0$ and $\tilde{l}_0$, $(e,e)=-1$.
The Poincare coordinates are: $h=(v,\tilde{l}_0)^{-1}$,
$x_1=(v,e)$, $x_2=(v, I.e)$, $x_0+x_3=(v,I.l_0)$ and
$x_0-x_3=(v,I.\tilde{l}_0)$.
\vspace{10pt}

\noindent
{\em General form of the extremal surface near the point of the boundary.}
We assume that the coordinates of the extremal surface have a series
expansion in $x_1$, $h$ near the point of the boundary $x_{\mu}=h=0$.
The equation for the extremal surface is to the lowest order in $h,x_1$:
\begin{equation}\label{MinimalLocally}
(\partial^2_{x_1}+\partial^2_h)y-{2\over h} \partial_h y=0
\end{equation}
where $y=x_0,x_2,x_3$.
If the contour near $x_{\mu}=h=0$ is a straight line plus corrections of
the order $x_1^3$ then the extremal surface is given near the point
$x_{\mu}=h=0$ by the expression cubic in $x_1$ and $h$ plus higher
orders. There are four cubic monomials $x_1^3, x_1^2 h, x_1 h^2, h^3$
and the equation (\ref{MinimalLocally}) leaves two combinations:
\begin{equation}\label{CubicMinimal}
y=\alpha h^3 + \beta (x_1^3  + 3 h^2 x_1)+\ldots
\end{equation}
where dots denote terms of the degree higher than 3 (the degree
of the monomial $h^a x_1^b$ is $a+b$.)
This is the general form of the extremal surface near the point
of the boundary where the curvature of the Wilson loop is zero.
The role of the coefficients $\alpha$ and $\beta$ is very different.
The coefficients $\beta$ are determined from the local behavior of 
the contour near the point $l_0$ (they measure the cubic deviation of the
contour from the straight line.) But
the coefficients $\alpha$ of $h^3$ depend globally on the contour.
They may be defined as the acceleration of the geodesic on the extremal
surface starting from the point $l_0$. To determine $\alpha$ for the
general contour we have to actually know the extremal surface. But
for the special contact contour we can determine the
direction of $\alpha$ without actually knowing the extremal surface.
\vspace{10pt}

\noindent
{\em The direction of $\alpha$.}
In the case of the special Legendrian surfaces, we
want to prove that $\alpha$ is directed along $I.l_0$.
Let us first consider the case when the contour is locally
a straight line. Near $x_1=0$, $h=0$ the surface should have the form:
\begin{equation}
\begin{array}{rl}
(z_0;z_1,z_2)=&{1\over h}
\left( {1+x_1^2+h^2\over 2}+i\alpha_0(x_1) h^3;\right.\\[5pt]&\left.
x_1+i\alpha_1(x_1) h^3,\;\; {i-ix_1^2-ih^2\over 2}+\alpha_2(x_1) h^3\right)
\end{array}
\end{equation}
Let us understand when this surface is Legendrian:
\begin{equation}\label{SLLeg}
\begin{array}{l}
\mbox{Im}\;\left[z_0^* \partial_{x_1} z_0 -
z_1^* \partial_{x_1} z_1 - z_2^* \partial_{x_1} z_2\right]=
\\[5pt]=
{1\over 2}
\left[(1+x_1^2)\partial_{x_1} \alpha_0
-2x_1\partial_{x_1}\alpha_1+
(1-x_1^2)\partial_{x_1}\alpha_2 -
2x_1\alpha_0+2\alpha_1+2x_1\alpha_2\right]h+\\[5pt]+o(h)=0
\\[10pt]
\mbox{Im}\;\left[z_0^* \partial_h z_0 -
z_1^* \partial_h z_1 - z_2^* \partial_h z_2\right]=
\\[5pt]=
{3\over 2}\left[(1+x_1^2)\alpha_0-2x_1\alpha_1+
(1-x_1^2)\alpha_2\right]+o(1)=0,\;\;\; h\to 0
\end{array}
\end{equation}
The condition for being special coincides with the second of these
equations:
\begin{equation}\label{SLSpec}
\mbox{Re}\; \iota_E\Omega= {3\over 2}
\left((1+x_1^2)\alpha_0-2x_1\;\alpha_1+(1-x_1^2)\alpha_2\right)
{dx_1\wedge dh\over h}+\ldots=0
\end{equation}
because all the extremal Legendrian manifolds
are special Legendrian.
From (\ref{SLLeg}) we find
\begin{equation}\label{ConditionOnAlpha}
\left[\begin{array}{r}
\alpha_0\\[5pt]
\alpha_1\\[5pt]
\alpha_2
\end{array}\right]=a(x_1)
\left[\begin{array}{c}
1+x_1^2\\[5pt]
2x_1\\[5pt]
-1+x_1^2
\end{array}\right]
\end{equation}
This means that $\alpha$ is directed along $I.l$ which is what we wanted
to prove.

Now suppose that the contour has a general cubic shape (\ref{CubicShape})
rather than being a straight line. 
The extremal surface has the following shape:
\begin{equation}
\begin{array}{l}
(z_0;z_1,z_2)={1\over h}
\left({1+x_1^2+h^2-\Phi_{\mu}^2\over 2}+i\Phi_0(x_1,h);
\right.\\[5pt]\left.
x_1+i\Phi_1(x_1,h),\;\; i{1-x_1^2-h^2+\Phi_{\mu}^2\over 2}+\Phi_2(x_1,h)
\right)
\end{array}
\end{equation}
where $\Phi_{\mu}^2=\Phi_0^2-\Phi_1^2-\Phi_2^2$. We assume that 
$\Phi(x_1,h)$ has a series expansion in $x_1$ and $h$ starting with cubic
terms:
$$
\Phi_{\mu}(x_1,h)=\alpha_{\mu} h^3 +\beta_{\mu}x_1^3 +\gamma_{\mu} x_1 h^2
+\delta_{\mu}x_1^2 h
+\ldots
$$
In fact we know that $\gamma_{\mu}=3\beta_{\mu}$ and $\delta_{\mu}=0$.
 The condition for being Legendrian is:
\begin{equation}
 (1+x_1^2+h^2-\Phi_{\mu}^2)\lrdiff \Phi_0
-2 x_1\lrdiff \Phi_1 +(1-x_1^2-h^2+\Phi_{\mu}^2)\lrdiff \Phi_2=0
\end{equation}
The term on the left hand side 
of the lowest degree in $x_1$ and $h$ comes from $d(\Phi_0+\Phi_2)$ and has
degree three. (We count $x_1, h, dx_1, dh$ as having the degree one.)  
This implies that $\Phi_0+\Phi_2$ does not have degree three terms (starts
with the degree four.) Therefore $\alpha_0+\alpha_2=0$.
The next term is quartic and it comes from 
$d(\Phi_0+\Phi_2)-2 x_1\lrdiff \Phi_1$. This implies that the quartic term
in $dx_1\wedge d\Phi_1$ is zero. Therefore $\alpha_1=\gamma_1=\delta_1=0$. 
But $\alpha_1=\alpha_0+\alpha_2=0$ means that $\alpha$ is directed
along $I.l_0$.

\section{Variation of the regularized area.}
The Wilson loop functional in the strong coupling regime is
the exponential of the regularized area of the extremal 
surface \cite{MaldacenaWilson}.
We should place the boundary of the extremal surface at 
small constant $h=h_0$ rather than $h=0$. The regularized
area of the surface is
\begin{equation}
A_{reg}=A-{1\over h_0}L[C]
\end{equation}
where $L[C]$ is the length of the contour $C$ in the
metric $dx_{\mu}^2$.
The variation of the loop can be described by the displacement
vector $\delta\;x^{\mu}$. The variation of the area  defines
the vector field $p_{\mu}(\sigma)$:
\begin{equation}
\delta A_{reg}[C]=\int_C dl\; p_{\mu}(\sigma)\;\delta x^{\mu}(\sigma)
\end{equation}
This vector field $p_{\mu}$ can be found from the shape
of the extremal surface near the boundary. 
Consider the  geodesic on the extremal surface
originating from the point on the boundary. The acceleration
of this geodesic (orthogonal to the extremal surface)
is $h^3p^{\mu}+o(h^3)$. The leading term does not depend
on the choice of the geodesic and gives a geometrical definition
of $p^{\mu}$. In particular, for the circular Wilson loop
$p^{\mu}=0$ --- the circular Wilson loop is an extremum
of the functional $W[C]$. In the Poincare coordinates near
the point $l_0$ of the boundary:
\begin{equation}
p^{\mu}=3\alpha^{\mu}
\end{equation}
where $\alpha^{\mu}$ is determined by (\ref{CubicMinimal}).

Let us explain why the variation of the regularized area is 
related to the acceleration of a geodesic near the
boundary. Consider the variation of the
extremal surface corresponding to the variation
of the contour on the boundary.  We can describe the variation of
the extremal surface by the normal vector $\xi(\sigma_1,\sigma_2)$.
The vector describing the variation of the surface is fixed
only up to vectors parallel to the surface; but
it is essential for our argument that we choose $\xi(\sigma_1,\sigma_2)$
to be normal to the surface. We need to regularize the area 
by choosing a boundary of the surface, for example by cutting
the surface at $h=h_0$ for small enough $h_0$. Since it is
a good regularization it does not matter how precisely we choose
the boundary. It is very natural to define the boundary
of the deformed surface to be the displacement of the boundary
of the original surface by the vector field $\xi$. Then the
variation of the area $A$ of the surface with the boundary will
be zero. Indeed, the variation of the area of the surface
is the integral over the surface of the trace of the second fundamental form
contracted with $\xi$. But for the extremal surface the trace
of the second fundamental form is zero. 
The only reason why $A_{reg}$ changes is the variation
of the length of the boundary which we subtract. The variation of
the length of the contour is the integral over the contour
of its acceleration contracted with $\xi$. But again, let us
take into account that the trace of the second fundamental
form of the extremal surface is zero. This means that
the normal component of the acceleration of the boundary 
is minus the normal component of the acceleration of
a curve orthogonal to the boundary, which we can choose
to be a geodesic on the surface.  

Calculation of $p^{\mu}(\sigma)$ or equivalently the acceleration
of the geodesic starting from the boundary requires the full 
knowledge of the extremal surface. There is no simple general
formula expressing the $p^{\mu}$ in terms of the contour $C$.
However in the special case of the special contact Wilson loop 
we know from the previous section that the acceleration is directed 
along the lightlike vector $I.l^{\mu}$.  
In Poincare coordinates:
\begin{equation}
\begin{array}{l}
{\delta\over\delta C_{\mu}(\sigma)}W[C]=
p^{\mu}(\sigma)=(p^0,p^1,p^2,p^3)= c(\sigma) \lambda^{\mu}(x(\sigma))
=\\[5pt]
=c(\sigma)\left({1+x_-^2+x_1^2+x_2^2\over 2},\;\;
-x_2+x_-x_1,\;\;x_1+x_-x_2,\;\;-{1+x_-^2-x_1^2-x_2^2\over 2}
\right)
\end{array}
\end{equation}
All the infinitesimal variations with $\delta x^{\mu}(\sigma)$
orthogonal for any $\sigma$ to $I.l(\sigma)$ will not change
the regularized area. 

For the example considered in Section 2.3 it turns out
that $p^{\mu}=0$, just as for the circular Wilson loop.
Indeed let us compute $p_{\mu}(\sigma=0)$. 
Let us choose the Poincare coordinates with the
origin at $\sigma=0$. Notice that $z_0, z_1, z_2$ are
odd functions of $\sqrt{T}$ when $T$ is large and $\sigma=0$. 
This means that $h$ is an odd function of $\sqrt{T}$ and 
$x_0,\ldots, x_3$ are even functions of $\sqrt{T}$. This means 
that  $x_{\mu}$ is an even function of $h$ and therefore 
it cannot have an  $h^3$ term, which is just what we wanted 
to prove. It would be interesting to understand in general 
which contours $C$ extremise $W[C]$.

\section*{Acknowledgements.}
I want to thank Yu.~Makeenko and K.~Zarembo for discussions
on the supersymmetry
of the circular Wilson loop.
This work was supported in part by the National Science
Foundation under Grant No. PHY99-07949 and in part
by the RFBR Grant No. 00-02-116477 and in part by the 
Russian Grant for the support of the scientific schools
No. 00-15-96557. 

\appendix

\section{Special Lagrangian manifolds are extrema of the area
functional.}
Consider a three dimensional submanifold $X\subset {\bf R}^{2+4}$.
Let us introduce on $X$ the coordinates $\sigma_{\mu}$, $\mu=0,1,2$.
The volume of $X$ is
\begin{equation}
\mbox{vol}\;X=\int_X d\sigma_0 \wedge d\sigma_1\wedge d\sigma_2
\sqrt{(w,w)}
\end{equation}
where $w={\partial\over\partial\sigma_0}\wedge
{\partial\over\partial\sigma_1}\wedge{\partial\over\partial\sigma_2}$
is the three-vector tangent to the surface. 
Suppose that ${\bf R}^{2+4}$ has a complex structure which makes
it ${\bf C}^{1+2}$.
Let us define $w^{3,0}$:
\begin{equation}
w^{3,0}=\left({1+iI\over 2}\otimes {1+iI\over 2}\otimes 
{1+iI\over 2}\right)\;.w
\end{equation}
The key point is that the space of $(3,0)$ forms has complex
dimension one. For that reason, there exists a $(3,0)$ form
$\Omega$ such that 
$(\overline{w^{3,0}},w^{3,0})=|\Omega(w^{3,0})|^2=|\Omega(w)|^2$.
On the other hand
\begin{equation}
(\overline{w^{3,0}},w^{3,0})=
(w,w)-3(w,(I\otimes I\otimes 1).w)
\end{equation}
Therefore the volume is:
\begin{equation}
\begin{array}{l}
\mbox{vol}\;X=\int_X d\sigma_0 \wedge d\sigma_1\wedge d\sigma_2
=\\[5pt]=
\int d\sigma_0 d\sigma_1 d\sigma_2 \sqrt{(\overline{w^{3,0}},
w^{3,0})
+3 (w, (I\otimes I\otimes 1).w)}
=\\[5pt]=
\int d\sigma_0 d\sigma_1 d\sigma_2 \sqrt{(\mbox{Re}\;\Omega(w))^2+
(\mbox{Im}\; \Omega(w))^2
+3 (w, (I\otimes I\otimes 1).w)}
=\\[5pt]=
\int \mbox{Re}\;\Omega +\ldots
\end{array}
\end{equation}
where dots denote the terms which are second order in the
variation of $X$ when $X$ is special Lagragian. (Notice
that $(w, (I\otimes I\otimes 1).w)$ is of the second order
in the deformation because $\left({\partial\over\partial \sigma^{\mu}},
I.{\partial\over\partial \sigma^{\nu}}\right)$ is of the first
order in the deformation.)
Since $\mbox{Re}\;\Omega$ is a closed form, 
this proves that $X$ extremises the volume functional.


\begin{thebibliography}{}
\bibitem{Maldacena}{J.M.~Maldacena, ``The Large N 
Limit of Superconformal Field Theories and
Supergravity'',
Adv.Theor.Math.Phys. 2 (1998) 231-252; 
Int.J.Theor.Phys. 38 (1999) 1113-1133;
hep-th/9711200.}
\bibitem{GKP}{S.S. Gubser, I.R. Klebanov, A.M.
Polyakov, 
``Gauge Theory Correlators from Non-Critical String
Theory'',
Phys.Lett. B428 (1998) 105-114, hep-th/9802109.}
\bibitem{Witten}{E.~Witten, ``Anti De Sitter Space And
Holography'',
Adv.Theor.Math.Phys. 2 (1998) 253-291,
hep-th/9802150.}
\bibitem{ReyYee}{S.J.~Rey, J.T.~Yee,
"Macroscopic strings as heavy quarks: Large-N gauge 
theory and anti-de Sitter supergravity",
Eur.Phys.J. C22 (2001) 379-394, hep-th/9803001.}
\bibitem{MaldacenaWilson}{J.M. Maldacena, 
"Wilson loops in large N field theories", 
Phys.Rev.Lett. 80 (1998) 4859-4862, hep-th/9803002.} 
\bibitem{ReyTheisenYee}{S.J.~Rey, S. Theisen,
 J.T.~Yee, 
"Wilson-Polyakov Loop at Finite Temperature in Large N 
Gauge Theory and Anti-de Sitter Supergravity",
Nucl.Phys. B527 (1998) 171-186, hep-th/9803135.}
\bibitem{BCFM}{D. Berenstein, R. Corrado, W. Fischler, 
J. Maldacena,  "The Operator Product Expansion for Wilson Loops 
and Surfaces in the Large N Limit",
     Phys.Rev. D59 (1999) 105023, hep-th/9809188}
\bibitem{DGO}{N. Drukker, D.J. Gross, H. Ooguri, 
"Wilson Loops and Minimal Surfaces",
    Phys.Rev. D60 (1999) 125006, hep-th/9904191.} 
\bibitem{ESZ}{J.K. Erickson, G.W. Semenoff, K. Zarembo, 
"Wilson Loops in N=4 Supersymmetric Yang--Mills Theory", 
Nucl.Phys. B582 (2000) 155-175, hep-th/0003055.}
\bibitem{DG}{N.~Drukker, D.J.~Gross, 
"An Exact Prediction of N=4 SUSYM Theory for String Theory",
J.Math.Phys. 42 (2001) 2896-2914, hep-th/0010274.}
\bibitem{Kruczenski}{M.~Kruczenski, 
"A note on twist two operators in N=4 SYM and Wilson loops in 
Minkowski signature", hep-th/0210115.}
\bibitem{Makeenko}{Yu.~Makeenko, 
"Light-Cone Wilson Loops and the String/Gauge Correspondence",
hep-th/0210256.}
\bibitem{CastroUrbano}{I.~Castro and F.~Urbano, "New examples of minimal
Lagrangian tori in the complex projective plane", 
Manuscripta Math. 85 (1994) 265-281; 
"On a minimal Lagrangian submanifold of ${\bf C}^n$ 
foliated by spheres", Michigan Math. J. 46 (1999) 71-82.}
\bibitem{Haskins}{M.~Haskins, "Special Lagrangian Cones",
math.DG/0005164}
\bibitem{MarkGross}{M.~Gross, "Examples of Special Lagrangian 
Fibrations",Symplectic geometry and mirror symmetry (Seoul, 2000), 
81--109, World Sci. Publishing, River Edge, NJ, 2001, 
math.AG/0012002}
\bibitem{Joyce}{D.~Joyce, "Special Lagrangian $m$-folds
in ${\bf C}^m$ with symmetries", math.DG/0008021; 
"Constructing special Lagrangian $m$-folds in ${\bf C}^m$
by evolving quadrics", Math. Ann. 320 (2001), no. 4, 757--797, 
math.DG/0008155;
"Evolution equations for special Lagrangian 3-folds
in ${\bf C}^3$", Ann. Global Anal. Geom. 20 (2001), no. 4, 345--403,
 math.DG/0010036}
\bibitem{BGK}{M. Bianchi, M.B. Green, S. Kovacs,
"Instanton corrections to circular Wilson loops in N=4 
Supersymmetric Yang-Mills", 
JHEP 0204 (2002) 040, hep-th/0202003.} 
\bibitem{HarveyLawson}{R.~Harvey and H.B.~Lawson, Jr., 
"Calibrated geometries", Acta Math. 148 (1982) 47-157.}
\bibitem{McLean}{R.C.~McLean, "Deformations of calibrated 
submanifolds", Comm. Anal. Geom. 6 (1998) no.4, 705-747.}
\bibitem{CH}{R.~Courant, D.~Hilbert, "Partial Differential
Equations", 1962}
\end{thebibliography}
\end{document}